# Effect of oxygen concentration on the structural and magnetic properties of $LaRh_{1/2}Mn_{1/2}O_3$ thin films

W.C. Sheets,[1] A. E. Smith,[2] M. A. Subramanian,[2] and W. Prellier[1,*]

[1]*Laboratoire CRISMAT, CNRS UMR 6508, ENSICAEN,*

*6 Bd. Maréchal Juin, F-14050 Caen, France*

[2]*Department of Chemistry, Oregon State University, Corvalis, OR 97331, USA*

**ABSTRACT**

Epitaxial $LaRh_{1/2}Mn_{1/2}O_3$ thin films have been grown on (001)-oriented $LaAlO_3$ and $SrTiO_3$ substrates using pulsed laser deposition. The optimized thin film samples are semiconducting and ferromagnetic with a Curie temperature close to 100 K, a coercive field of 1200 Oe, and a saturation magnetization of 1.7 $\mu_B$ per formula unit. The surface texture, structural, electrical, and magnetic properties of the $LaRh_{1/2}Mn_{1/2}O_3$ films was examined as a function of the oxygen concentration during deposition. While an elevated oxygen concentration yields thin films with optimal magnetic properties, slightly lower oxygen concentrations result in films with improved texture and crystallinity.

[*] prellier@ensicaen.fr

## I. INTRODUCTION

The mixed *B*-site perovskites of composition $LaB_{1/2}Mn_{1/2}O_3$ have received renewed attention recently owing to their coexisting ferromagnetic and semiconducting/insulating properties.[1-15] Remarkable near room temperature ferromagnetic transition temperatures of 280 K and 230 K have been reported for both bulk and thin film samples of $La_2NiMnO_6$ (LNMO)[1,16] and $La_2CoMnO_6$ (LCMO)[2,17] perovskites, respectively. In thin film form, such ferromagnetic semiconductors have potential applications in next-generation spintronic devices, including spin-based transistors and advanced magnetic memory storage elements.[18] The structural and valence ordering of the transition metal cations on the *B*-site plays a vital role in determining the magnetic properties of the samples. In ordered samples it is expected that the empty $e_g$ orbitals of $Mn^{4+}$ interacts with the half-filled $e_g$ orbitals of $B^{2+}$ ($B$ = Co, Ni) through a ~180º $Mn^{4+}$-O-$B^{2+}$ Goodenough-Kanamori interaction, resulting in a ferromagnetic coupling. A decrease in order on the *B*-site, however, reduces the magnetic phase transition temperature by increasing the number of antiferromagnetic interactions. The composition of the *B*-site cations in $LaB_{1/2}B'_{1/2}O_3$ perovskites is not limited to third row transition metals, and, indeed, studies of bulk $LaRh_{1-x}B_xO_3$[19-23] and $LaIr_{1-x}B_xO_3$ ($B$ = Mn, Fe, or Ni)[24,25] samples have been reported previously. While high-spin configurations are anticipated for first row (3*d*) transition metals, a low-spin configuration is preferred for second (4*d*) and third row (5*d*) transition metals, owing to the increased splitting energy between the $e_g$ and $t_{2g}$ orbitals. In order to obtain a strong ferromagnetic virtual spin-spin superexchange interaction with heavier transition metal cations, such as $Rh^{4+}$ or $Ir^{4+}$ possessing empty $e_g$ orbitals, a complimentary high-spin first row transition cation with a partially-filled $e_g$ orbital is required on the *B*-site (e.g., $Mn^{2+}$, $Fe^{2+}$, or $Ni^{2+}$). In this letter, we investigate the thin film deposition of $LaRh_{1/2}Mn_{1/2}O_3$ (LRMO) and examine how its structure, morphology, and magnetic properties are altered by adjusting the oxygen background pressure during deposition.

Numerous studies have examined the structural and magnetic properties of $LaB_{1/2}Mn_{1/2}O_3$ samples to understand their local ordering on the *B*-site sublattice, or lack thereof. In particular, two different types of ordering are possible: the atomic (i.e., rocksalt, random, or a mixture of each) and oxidation state order of the transition metals (i.e., $B^{2+}/Mn^{4+}$ or $B^{3+}/Mn^{3+}$). Although a degree of mixed valency always can be anticipated, there has been disagreement over the predominant cation oxidation states. For example, three independent neutron diffraction studies report different manganese and nickel oxidation states in LNMO samples. Blasco et al. and Rogado et al. report the presence of $Ni^{2+}$ and $Mn^{4+}$ cations,[16,26] whereas Bull and coworkers conclude that $Ni^{3+}$ and $Mn^{3+}$ are present.[27] More recently, data from electron energy loss spectroscopy measurements, Raman scattering, and X-ray photoelectron spectroscopy indicate $Ni^{2+}$ and $Mn^{4+}$ to be the prevalent oxidation states in LNMO thin films.[14] However, the authors also observed an increase in charge disproportionation of $Ni^{2+}$ and $Mn^{4+}$ to $Ni^{3+}$ and $Mn^{3+}$ when decreasing the oxygen background pressure during film deposition. Similar to LNMO, an ideal $Rh^{4+}/Mn^{2+}$ charge distribution cannot be assumed for LRMO films, owing to the similar stabilities of tetravalent ($Rh^{4+}$) and trivalent ($Rh^{3+}$) rhodium oxidation states. Schnizer proposed an $Rh^{4+}/Mn^{2+}$ charge distribution because no structural distortion from $Mn^{3+}$, a $3d^4$ high-spin ion exhibiting a strong Jahn-Teller effect, was evident in the Rietveld refinement.[23] On the other hand, Haque and Kamegashira suggest $Rh^{3+}/Mn^{3+}$ valence states since the observed effective magnetic moments derived from the magnetic states of $LaRh_{1/2}Mn_{1/2}O_3$ in the paramagnetic region are near to the spin-only values of this combination.

In terms of the atomic order, the majority of studies conclude that a large amount of local rocksalt ordering occurs on the B-site sublattice, but not across a long-range scale. Indeed, all of the aforementioned neutron diffraction studies of bulk LNMO provide evidence for a locally ordered Ni/Mn atomic arrangement.[16,26,27] Refinements of these neutron diffraction studies, however, also suggest the presence of anti-site defects (i.e., Ni and Mn atoms are not ordered perfectly) within the ordered B-site sublattice, which when accounted for improve the refinement of the data. Data from polarized Raman spectra and measurement of saturation magnetization values near the theoretical maxima in LCMO and LNMO thin films corroborate that a large amount of cation ordering exists in such samples.[1,2,11,28] On a smaller scale, selected area electron diffraction reveals a majority *I*-centered phase coexisting with domains of a minority *P*-type phase, the latter which is dispersed throughout the *I*-type matrix.[7,14] The presence of the mirror plane *a* in the *I*-type phase and the numerous orientations observed for the *P*-type nano-domains argues against the complete long-range ordering of the Ni/Mn sublattice. In bulk $LaRh_{1/2}Mn_{1/2}O_3$ samples, no evidence for long-range ordering was observed in Rietveld refinements of x-ray diffraction patterns.[23] However, their magnetic properties indicate at least a partially ordered arrangement of the Rh/Mn cations exists, enabling ferromagnetic spin-spin superexchange interactions.

## II. THIN FILM DEPOSITION

Although samples of LRMO have been examined in the bulk, thin film samples have not been fabricated or studied. Multiple thin films were grown on both (001)-oriented $LaAlO_3$ (LAO) and $SrTiO_3$ (STO) substrates. Stoichiometric $LaRh_{1/2}Mn_{1/2}O_3$ was employed as a target, which was synthesized by conventional solid state methods. The films were deposited between 650–750 °C by the PLD technique using a KrF excimer laser (248 nm, 3 Hz) at different (10–800 mTorr) atmospheres of flowing oxygen under dynamic vacuum. On average, 5000 pulses yielded films with a thickness of 100–150 nm, depending on the oxygen background pressure. The crystalline structure of the thin film samples was examined by x-ray diffraction (XRD) using a Seifert 3000P diffractometer (Cu $K_{\alpha}$, $\lambda$ = 1.5406 Å). A Philips X'Pert diffractometer was used for the in-plane XRD measurements of the film samples. Figure 1a shows the XRD $\theta$-$2\theta$ pattern for a LRMO film grown on LAO at 720 °C and 300 mTorr oxygen pressure. The peaks were indexed based on a pseudo-cubic unit cell and only the peaks corresponding (00*l*) reflections (where $l$ = 1, 2, 3…) were observed, which indicates that the out-of-plane lattice parameters is a multiple of the perovskite subcell parameter ($a_{sub}$ = 3.93 Å). The absence of diffraction peaks from secondary phases or randomly orientated grains evidences the preferential orientation of the films. The full-widths-at-half-maximum (FWHM) is 0.072° for the (002) reflection of LRMO (inset of Fig. 1), as measured by XRD rocking-curve analysis ($\omega$-scan), and when compared with the FWHM of 0.06° measured for the $LaAlO_3$ substrate evidences the high-quality of the thin film sample. The in-plane orientation, as evaluated by the XRD $\Phi$-scan of the LRMO (103) reflection of the cubic subcell (Fig. 1b), shows four peak separated by 90° revealing the fourfold symmetry and indicating that the LRMO film is epitaxial with respect to the substrate. A large degree of in-plane texture $FWHM_{\Phi}$ =0.9° is observed for the films grown on LAO(001).

The out-of-plane lattice parameter $c$ and defect structure of LRMO films on STO(001) and LAO(001) substrates varies as a function of the oxygen pressure present during deposition. As shown in Figure 2, films deposited at lower oxygen pressure have an expanded lattice parameter $c$ when compared to the pseudocubic bulk value (3.93 Å), and increasing the oxygen pressure steadily decreases the lattice parameter $c$, which eventually becomes less than that of the bulk value. Previous studies have attributed this lattice expansion at lower oxygen pressures to an increase in the defect concentration.[12,14] Two mechanisms have been proposed to account for such an increase at lower oxygen pressures, i) oxygen vacancies caused by the lack of oxygen present during deposition and ii) lattice damage resulting from the increase in flux and energy of species colliding with the surface of the growing film. Studies comparing the growth of LNMO films in 50 mTorr oxygen and a 50 mTorr $Ar/O_2$ (10/1 ratio) mixture indicate that the latter defect type dominates.[14] The increased out-of-plane lattice parameter *c* observed in the present study most likely have the same origin – films grown at lower oxygen pressure having a greater defect concentration and lattice disorder owing to the increased surface bombardment by high energy species during growth. As summarized previously, this defect structure is especially important in mixed *B*-site perovskite films owing to the high degree of lattice order required to optimize their magnetic properties.

## III. RESULTS AND DISCUSSION

Scanning electron microscopy (SEM) and atomic force microscopy (AFM) images confirm the successful manipulation of LRMO film microstructure by altering the oxygen pressure during deposition (Figure 3). Two samples were selected based upon either their optimized crystallinity (300 mTorr) or magnetic properties (800 mTorr). As shown in Figure 3b, SEM images of LRMO films grown at higher oxygen pressure (800 mTorr) reveal the films are composed of relatively small, discrete grains. A layer of particles, which appear to be clusters of nanoparticles ~100 nm in diameter, is clearly visible with a single grain overgrowth. In contrast, at lower oxygen pressures (300 mTorr) the underlying films appear smooth and featureless by SEM (i.e., no observable grains at the maximum magnification). Small particulates remain on the surface (20-50 nm), but they are smaller in diameter and cover less of the film surface (Fig 3a). AFM images also reveal a distinct difference in the morphology between these two types of films (Figs. 3c &

d), albeit at smaller length scales. In addition to the visible difference in surface morphology, differences exist in the root mean square roughness ($R_{rms}$) and the maximum peak-to-valley roughness ($R_{pv}$) values. Films grown at lower oxygen concentration are smoother ($R_{rms}$ = 10 nm), although as can be observed in the SEM images numerous granules up to 100 nm in height ($R_{pv}$ = 104 nm) are scattered across the surface of the film. At higher oxygen growth pressures, AFM images indicate significant roughening has occurred, although a significant increase in root mean square roughness is not observed ($R_{rms}$ = 12 nm). The granular surface of LRMO thin films grown at 800 mTorr is better evidenced by the enhanced maximum peak-to-valley roughness ($R_{pv}$ = 192 nm) measured for these films.

Measurements of magnetization ($M$) versus applied magnetic field ($H$) and temperature ($T$) were performed on all samples using a superconducting quantum interference device magnetometer (SQUID). The zero-field cooled (ZFC) and field-cooled (FC) response for LRMO films deposited under different oxygen pressure was measured under a low applied field of 500 Oe and at a higher applied field of 10 kOe. Figure 4 illustrates the ZFC and FC magnetization curves measured at 500 Oe for samples deposited at 100, 300, 600, and 800 mTorr oxygen background pressure. The values of the ferromagnetic Curie temperature ($T_C$) were estimated from the minimum of the temperature derivative of the magnetization ($\delta M/\delta T$). Two different ferromagnetic Curie temperatures are observed in the series of LRMO films, and the presence, absence, or coexistence of each ferromagnetic phase is dependent on the oxygen background pressure during film growth. Samples deposited at high oxygen pressures (≥600 mTorr) display a single magnetic transition with a Curie temperature ($T_C$) of ~80 K, which corresponds well with the $T_C$ value of 72 K reported previously for polycrystalline LRMO samples.[23] In contrast, films deposited at low oxygen background pressures (≤ 100 mTorr) demonstrate a much lower $T_C$ ferromagnetic phase ($T_C \sim 30$ K). Films deposited at intermediate oxygen background pressures (e.g. 300 mTorr in Figure 4b) contain a mixture of the two ferromagnetic phases, evidenced by two different $\delta M/\delta T$ minimums – one corresponding to the high $T_C$ ferromagnetic phase ($T_C \sim 80$ K) and the other with the lower $T_C$ ferromagnetic phase ($T_C \sim 30$ K). The presence of two different ferromagnetic phases also has been observed in LCMO thin film samples, and also depends on the deposition conditions.[2] In particular, the lower $T_C$ ferromagnetic phase in LCMO films is associated with the existence of oxygen vacancies.[12] For all samples, the magnetization in the FC curves increase consistently with decreasing temperature, whereas the ZFC curves demonstrate a cusp shape below the ferromagnetic $T_C$. These cusps disappear when the magnetic field is increased to 10 kOe (inset of Figure 5), which is a characteristic of spin-glass behavior.[2] Application of a stronger magnetic field also increases the $T_C$ of the LRMO film to 104 K as the enhanced field suppresses the contribution from antiferromagnetic interactions.[23]

The magnetic hysteresis in-plane loops measured at 10 K for LRMO films grown at certain oxygen background pressures (100, 300, 600, and 800 mTorr) is shown in Figure 5. Both the saturation magnetization and coercive fields increase with rising oxygen background pressure, and the maximum value of saturation magnetization measured at 10 K for a LRMO film deposited at 800 mTorr is 1.7 $\mu_B$/f.u. The loop for this optimized sample shows a well-defined hysteresis with a coercive field approximately equal to 1.2 kOe, and a saturation field close to 5 kOe. The enhancement in the saturation magnetization with increasing background oxygen pressure likely results from an increased number and size of domains with ordering of the *B*-site cations and a reduction of oxygen vacancies.[12,14] A similar enhancement in the coercivity can be attributed to the accompanied reduction in magnetic wall domains within the thin film sample.

The magnetic properties of each potential cation combination between the *d* orbitals of high-spin $Mn^{2+}(3d^5)$ and low-spin $Rh^{4+}(4d^5)$ cations, through the oxygen 2*p* orbitals, can be interpreted based on the rules for the sign of spin-spin superexchange interactions,[29,30] and are summarized in Table I. Virtual spin transfer between half-filled to empty orbitals, $Mn^{2+}(e^2)$–O–$Rh^{4+}(e^0)$, along with full to half-filled orbitals, $Mn^{2+}(t^2)$ –O–$Rh^{4+}(t^4)$, yield ferromagnetic coupling. In contrast, antiferromagnetic coupling results from virtual spin transfer between half-filled to half-filled orbitals, for example, $Mn^{2+}(e^2)$–O–$Mn^{2+}(e^2)$ and $Rh^{4+}(t^1)$–O–$Rh^{4+}(t^1)$. Ferromagnetic interactions are stronger between e orbitals than t orbitals because the overlap of the σ-bonding e electrons is greater and their ΔE is smaller than that of the π-bonding electrons for t orbitals. Therefore, a well ordered LRMO film dominated by σ-bonding $Mn^{2+}(e^2)$–O–$Rh^{4+}(e^0)$ interactions is expected to generate ferromagnetic coupling. An increase in point disorder, however, enhances the number of $Mn^{2+}(e^2)$–O–$Mn^{2+}(e^2)$ and $Rh^{4+}(t^1)$–O–$Rh^{4+}(t^1)$ antiferromagnetic interactions within the sample. In addition, antiphase grain boundaries, where the cation positions are inverted, can also yield a significant number of such antiferromagnetic interactions. Competition between these ferromagnetic and antiferromagnetic interactions in partially B-site disordered samples leads to complex magnetic behavior and most likely produces the observed spin-glass behavior in LRMO films.[12] The alternative high-spin Jahn-Teller $Mn^{3+}(3d^4)$ and low-spin $Rh^{3+}(4d^6)$ cation combination must also be considered. An ordered $Rh^{3+}/Mn^{3+}$ interaction generates both σ-bonding $Mn^{3+}(e^1)$–O–$Rh^{3+}(e^0)$ and π-bonding $Mn^{3+}(t^3)$ –O–$Rh^{3+}(t^6)$ ferromagnetic interactions. In the case of disorder, however, three-dimensional ferromagnetic vibronic superexchange between Jahn-Teller cations $Mn^{3+}(e^1)$–O–$Mn^{3+}(e^1)$ and diamagnetic $Rh^{3+}(t^6)$–O–$Rh^{3+}(t^6)$ interactions are also possible.[31] Such vibronic superexchange between $Mn^{3+}(e^1)$–O–$Mn^{3+}(e^1)$ provides less stabilization when compared to the ferromagnetic superexchange of $Mn^{2+}$–O–$Rh^{4+}$ cations, and may account for the lower $T_C$ ferromagnetic phase.[12] At the same time, the observation of an additional low temperature maximum in the AC susceptibility of polycrystalline LRMO has been attributed to either i) competing interactions antiferromagnetic and ferromagnetic interactions between identical and different species of neighboring cations in statistically ordered domains or ii) ferromagnetic domains coexisting with an important paramagnetic volume fraction.[23] Nonetheless, an increase in oxygen vacancies reduces the number of ferromagnetic interactions from ordered $Mn^{2+}$–O–$Rh^{4+}$ lattices, and possibly explains why the contribution from the higher $T_C$ ferromagnetic phase is diminished significantly in films deposited at lower oxygen background pressure.

The maximum value of saturation magnetization (1.7 $\mu_B$/f.u.) is below the expected spin-only value ($\mu_{calc}$) of 4.36 $\mu_B$/f.u for $Rh^{4+}$/$Mn^{2+}$ valence states[23] or 4.65 $\mu_B$/f.u for $Rh^{3+}$/$Mn^{3+}$ valence states.[21] For polycrystalline LRMO samples, using the former spin-only value, the magnetic data indicated the ferromagnetic volume fraction to be ~35 %, which can be attributed to the fraction of cation ordered domains, and therefore indicating that the majority of the B-site sublattice remains disordered. A similar calculation comparing the saturation magnetization of the optimized LRMO films at 10 K with the spin-only value yields a slightly enhanced ferromagnetic volume fraction of ~40 %. Films grown at lower oxygen concentrations possess reduced ferromagnetic volume fractions with the minimum value reaching 7% for a LRMO thin film grown at 10 mTorr oxygen background pressure. These results are consistent with an increase in structural defects, and therefore antiferromagnetic superexchange interactions, in thin films grown at lower oxygen pressure.

The dc-electrical properties of the film samples were measured by a physical property measurement system (PPMS) in four-probe configuration. Figure 6 illustrates the resistivity as a function of temperature. As expected, the thin film samples were semi-conducting with a high resistivity of ~$10^1$ Ω cm at room temperature, which increases with decreasing temperature until the resistance exceeds the maximum value that can be measured ($R > 10^7$ Ω) at ~200 K. An Arrenhius plot of the log $\rho$ versus inverse temperature (inset of Figure 6) reveals a linear dependence, indicating semiconductor behavior with a low activation energy of 0.16 eV.

Magnetodielectric responses have been reported previously for ferromagnetic $LaB_{1/2}Mn_{1/2}O_3$ polycrystalline and thin film samples.[13,16] Epitaxial bilayer LRMO/LNO films were grown on (0010-oriented LAO substrates to examine the dielectric response of LRMO under various temperatures, frequencies, and magnetic fields. Prior to the deposition of 100 nm thick LRMO film, a bottom electrode of $LaNiO_3$ (50 nm) was deposited at 700 °C at an oxygen pressure of 100 mTorr. Square gold pads of 400 $\mu$m x 400 $\mu$m dimensions (physical mask) were sputtered on top of the LRMO films and the LNO regions unexposed to the LRMO deposition. The fabricated heterostructures were characterized in the metal-insulator-metal configuration to study their dielectric properties. Figure 7 shows the temperature dependence of the dielectric permittivity ($\varepsilon$) at selected frequencies for a LRMO film deposited with an oxygen back pressure of 800 mTorr. At low temperature the dielectric permittivity at all frequencies remains near 10-20 until $\varepsilon$ increases to a maximum at temperatures near 150-175 K. Then, the dielectric permittivity decreases to a minimum at higher temperature where values near 1 are reached. There is large frequency dispersion as illustrated by both the decrease in intensity and shift to higher temperatures for the maximum value of the dielectric permittivity with increasing applied frequency. Although the dielectric anomaly occurs near temperatures observed for the magnetic ordering temperature of LRMO, such enhancements in the capacitance of a semiconductor with observable grain boundaries likely are extrinsic.[32,33] As summarized by Scott,[34] Maxwell-Wagner effects can occur at grain boundaries near the magnetic ordering temperature of magnetoresistive materials, and can lead to enhancements of the dielectric constant exceeding 1000%. Indeed, our plot matches closely with the calculated real part of the dielectric constant in a Maxwell-Wagner equivalent circuit reported previously, suggesting such artifacts predominate.[34] As shown in Figure 8, the films are lossy and the dielectric loss remains relatively constant at all temperatures, data which provide further evidence that the observed dielectric anomaly is unrelated to true magnetoelectric coupling. It should also be noted that unlike polycrystalline and thin film samples of the double perovskite LNMO, no significant change in the dielectric permittivity of LRMO films was observed upon the application of magnetic fields up to 10 kOe at a constant frequency (50 kHz).

## IV. CONCLUSIONS

In conclusion, epitaxial LRMO thin films have been deposited on both (001)-oriented STO and LAO using the pulsed laser deposition technique. A change in the defect structure, surface morphology, and magnetic properties was observed for LRMO films grown at different oxygen pressures. All films are semiconducting and ferromagnetic; however, the Curie temperature, coercive field, and saturation magnetization of LRMO films are dependent on the oxygen background pressure during deposition. Higher oxygen deposition pressures (>600 mTorr) yield films with a rougher surface morphology and a reduced defect structure, the latter which improves the local ordering of the B-site cations and enhances the magnetization of the film. Films deposited at lower oxygen pressures display a second low temperature $T_C$ ferromagnetic phase, which may be associated with the alternate $Mn^{3+}/Rh^{3+}$ valence state. Such results further demonstrate the critical role oxygen background pressure during film growth plays in controlling the morphology, crystalline structure, and magnetic properties of $LaB_{1/2}Mn_{1/2}O_3$ perovskite films.

**ACKNOWLEDGMENTS**

The authors thank R. Ranjith, B. Kundys, L. Goulef, J. Aubril, and Y. Thimont for their helpful discussions and assistance in film preparation and sample characterization. This work is carried out in the frame of the NoE FAME (FP6-5001159-1), the STREP MaCoMuFi (NMP3-CT-2006-033221) and the STREP CoMePhS (NMP4-CT-2005-517039) supported by the European Community and by the CNRS, France. Partial support from the ANR (NT05-1-45177, NT05-3-41793) is acknowledged. The work performed at the Oregon State University is supported by the grant from National Science Foundation (DMR 0804167). WCS was additionally supported through a Chateaubriand post-doctoral fellowship and the Conseil Régional de Basse-Normandie (CRBN). MAS thanks the CRBN for supporting a Chair of Excellence at Laboratoire CRISMAT.

**Table I.** Expected super-exchange interactions for ordered LRMO films

| B-cation | | B'-cation | | Interaction | Result |
|---|---|---|---|---|---|
| $Rh^{4+}$, $4d^5$ (LS) | | $Mn^{2+}$, $3d^5$ (HS) | | | |
| 2 $e_g$ | Empty | 2 $e_g$ | Half | σ | FM |
| 1 $t_{2g}$ | Half | 1 $t_{2g}$ | Half | π | AFM |
| 2 $t_{2g}$ | Full | 2 $t_{2g}$ | Half | π | FM |
| | | | | | |
| $Rh^{4+}$, $4d^5$ (LS) | | $Rh^{4+}$, $4d^5$ (LS) | | | |
| 1 $t_{2g}$ | Half | 1 $t_{2g}$ | Half | π | AFM |
| | | | | | |
| $Mn^{2+}$, $3d^5$ (HS) | | $Mn^{2+}$, $3d^5$ (HS) | | | |
| 2 $e_g$ | Half | 2 $e_g$ | Half | σ | AFM |
| 3 $t_{2g}$ | Half | 3 $t_{2g}$ | Half | π | AFM |

## FIGURE CAPTIONS

Figure 1. A XRD $\theta$-$2\theta$ scan curve of a typical LRMO film grown on LAO substrate. The substrate ($00l$) are marked in the figures. The inset shows a rocking curve recorded around the 001 reflection of the film.

Figure 2. A plot of the out-of-plane $c$-axis lattice parameter of LRMO thin films grown on STO and LAO substrates as a function of the oxygen background pressure during growth.

Figure 3. (color online). SEM images of LRMO films grown at a) 300 mTorr and b) 800 mTorr along with three-dimensional AFM images showing the surface of LRMO films deposited on LAO at c) 300 mTorr and d) 800 mTorr oxygen background pressure. The scan size is 500 nm by 500 nm.

Figure 4. A plot of the temperature dependence of the ZFC and FC responses of LRMO films grown under different oxygen background pressures of a) 100 mTorr, b) 300 mTorr, c) 600 mTorr, and d) 800 mTorr.

Figure 5. (color online). A plot of M(H) hysteresis loops for LRMO films grown under different oxygen pressures on LAO substrates measured at 10 K. The inset displays the M(T) curve recorded with 1 T. Magnetic field is applied parallel to the [100] direction of the substrate.

Figure 6. A plot of the resistivity of LRMO films as a function of temperature. The inset displays the log $\rho$ as a function of inverse temperature (Arrhenius plot).

Figure 7. (color online). A plot of the temperature dependence of the real part of the dielectric response ($\varepsilon$) measured at certain frequencies for a bilayer LRMO/LNO film deposited on (001)-oriented LAO.

Figure 8 (color online). A plot of the dielectric loss as a function of temperature measured at certain frequencies.

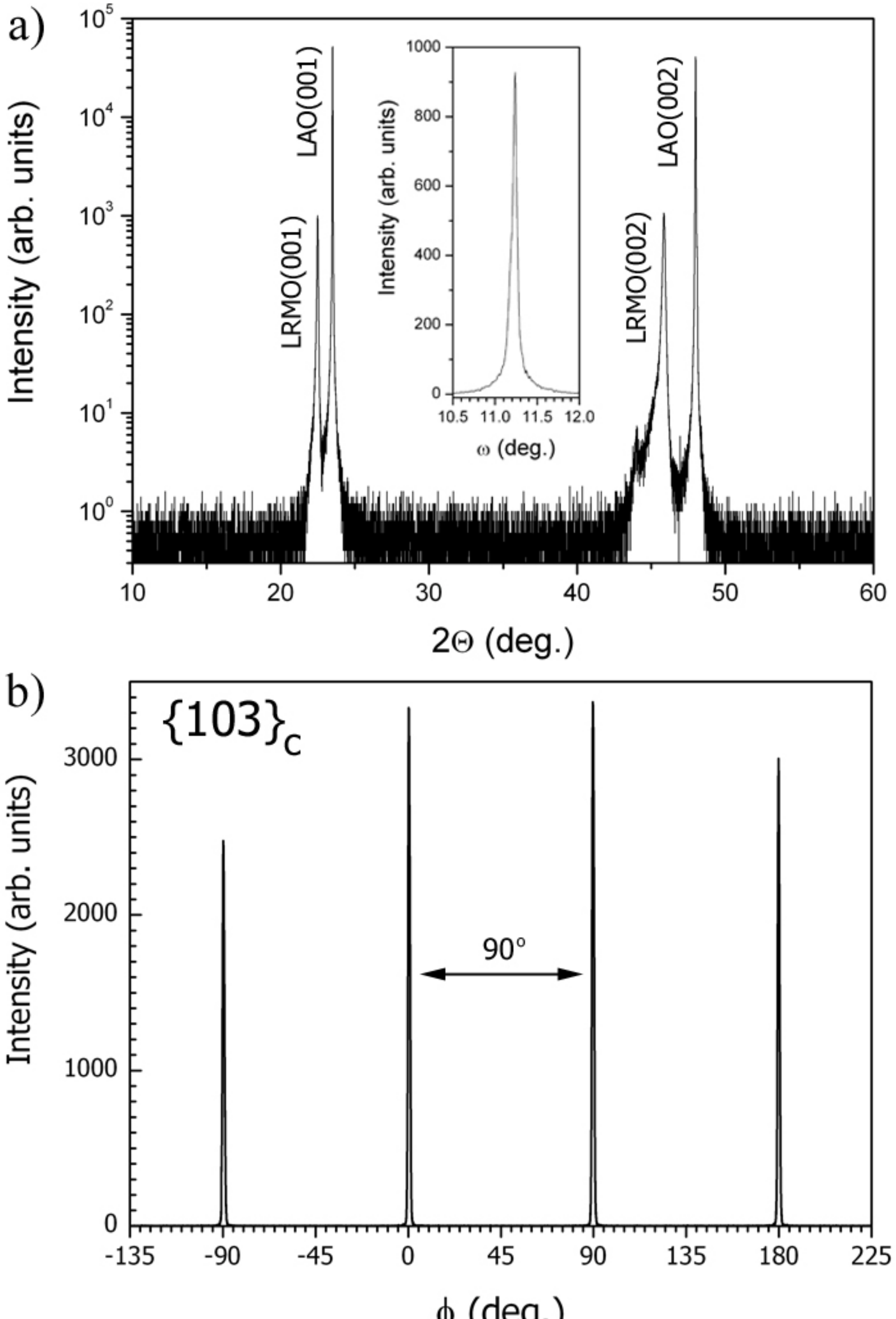

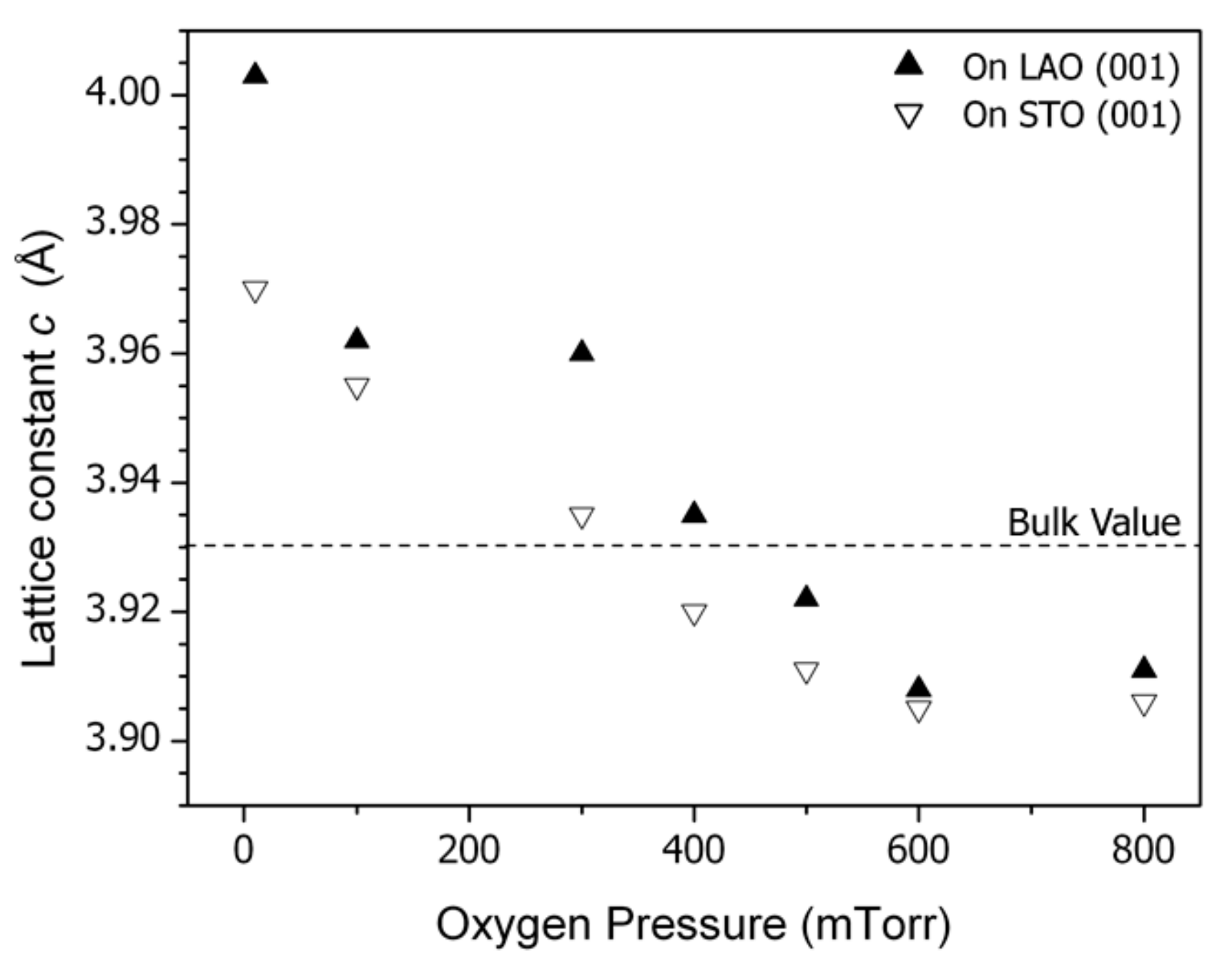

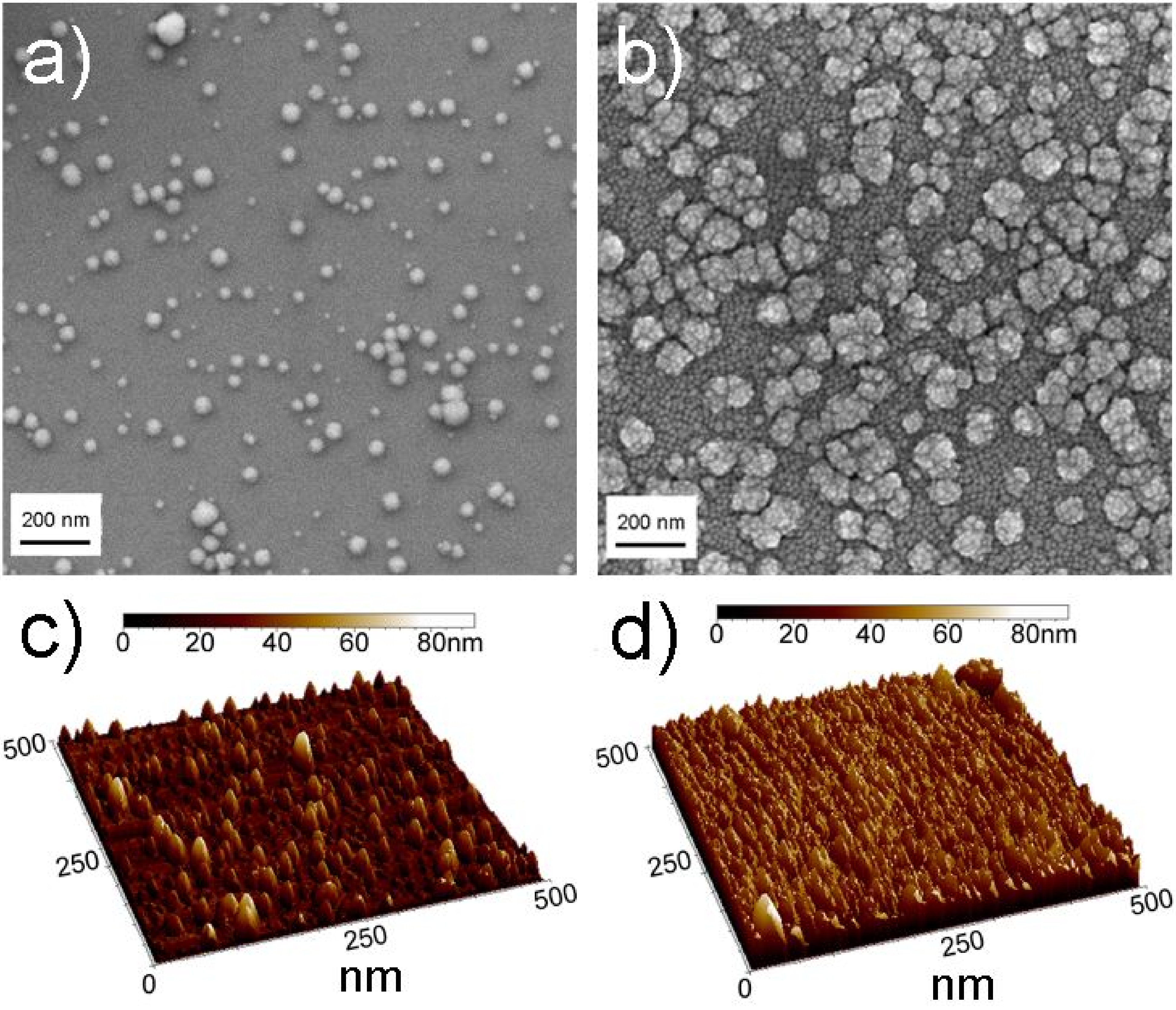

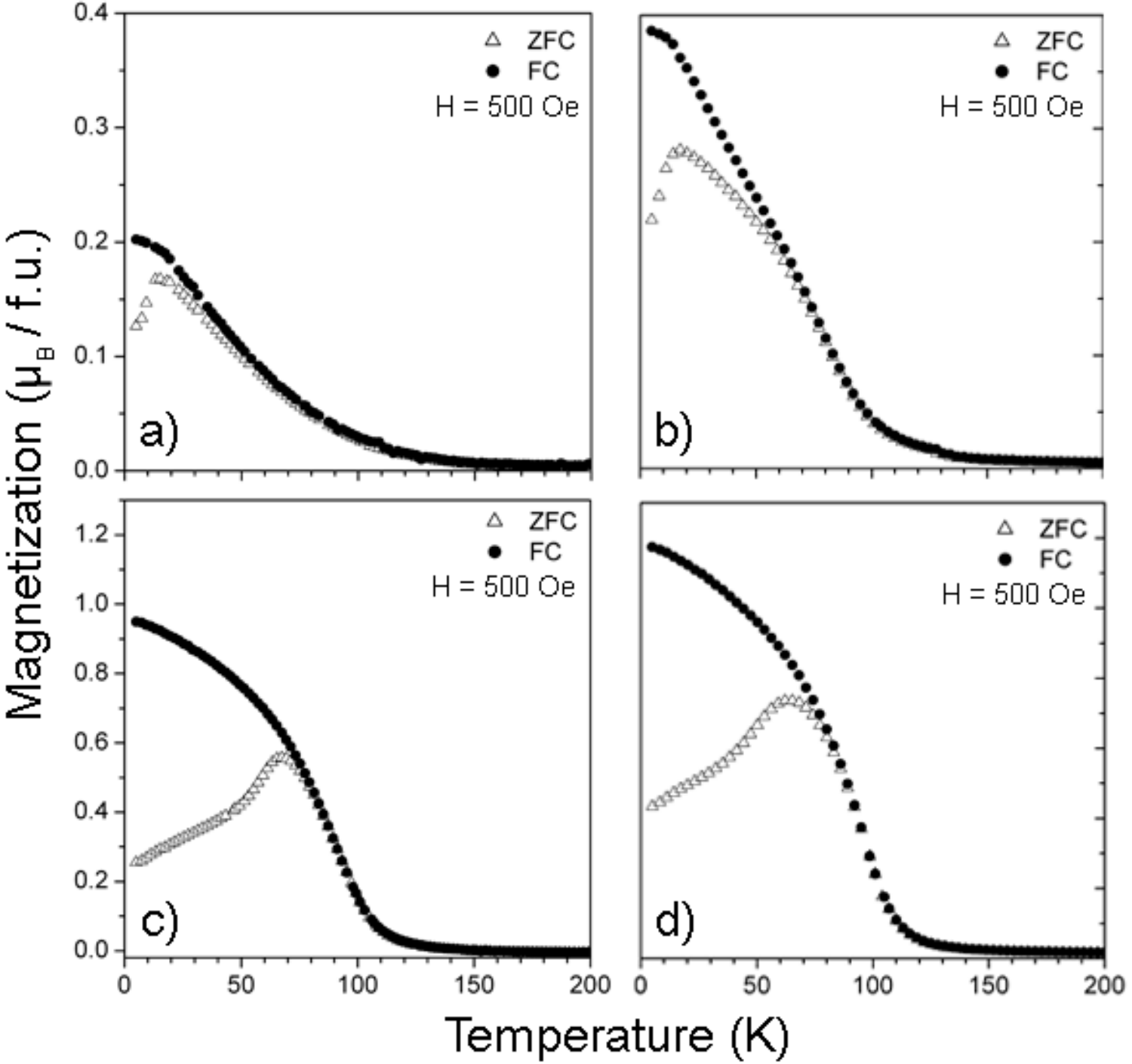

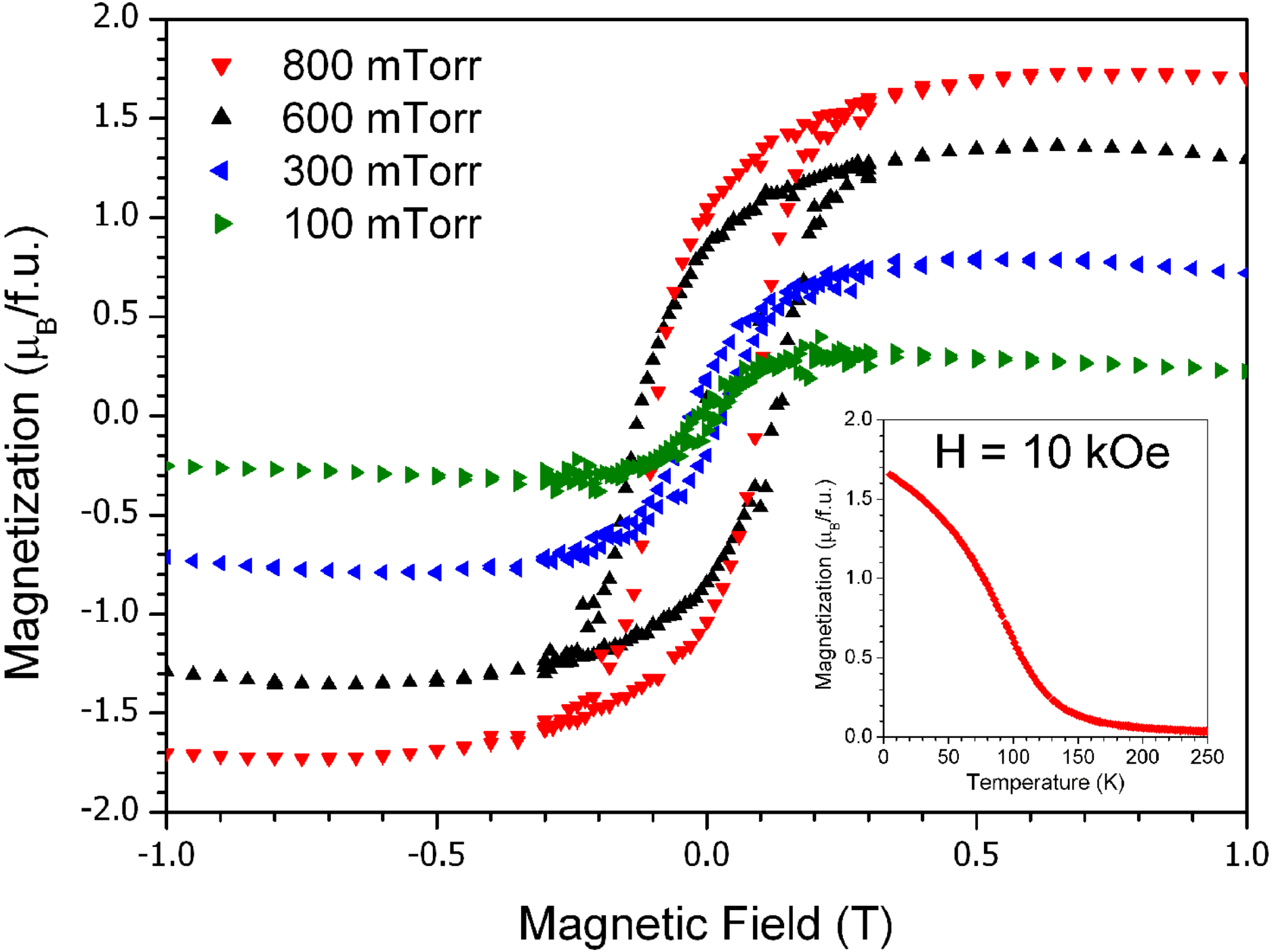

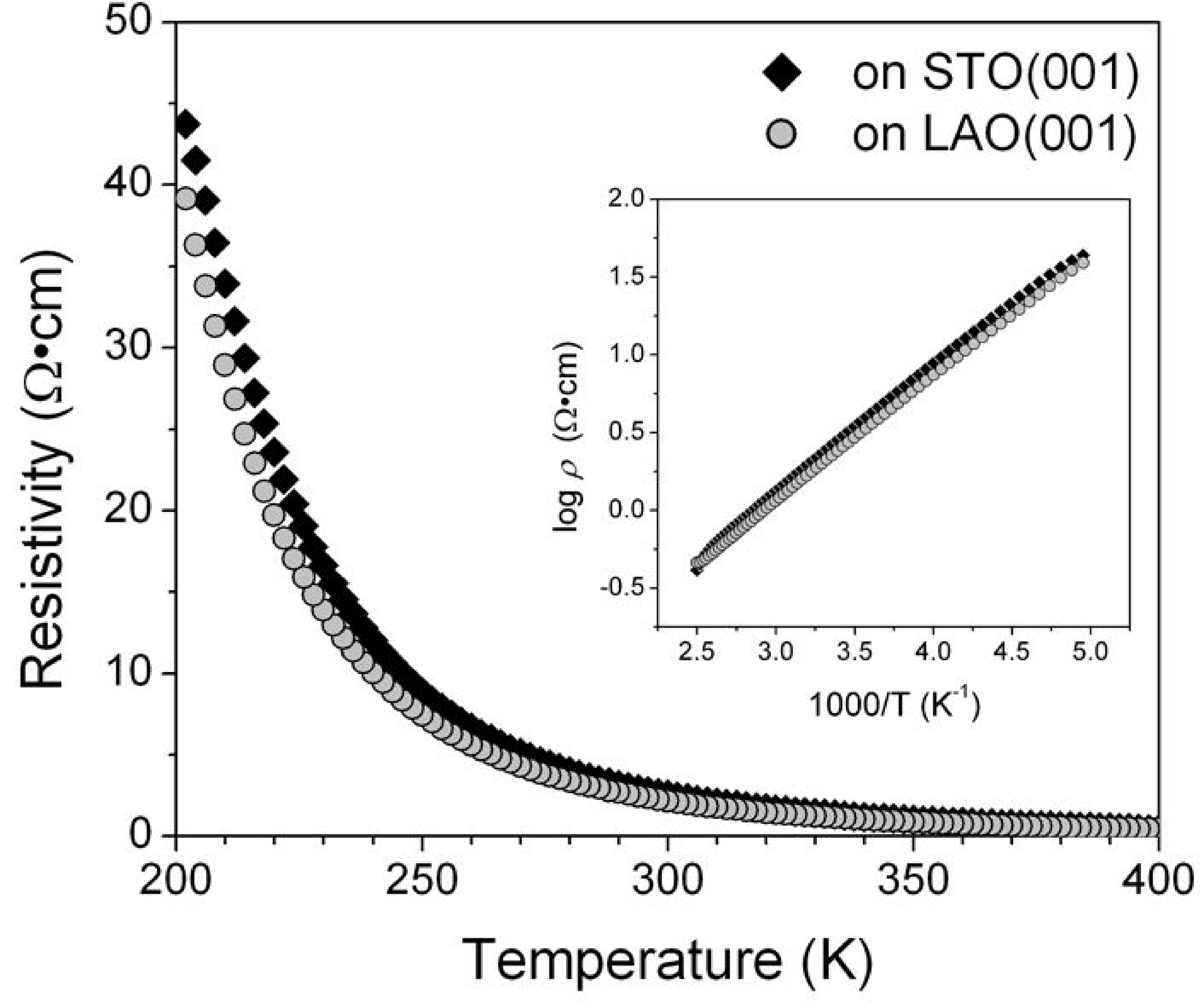

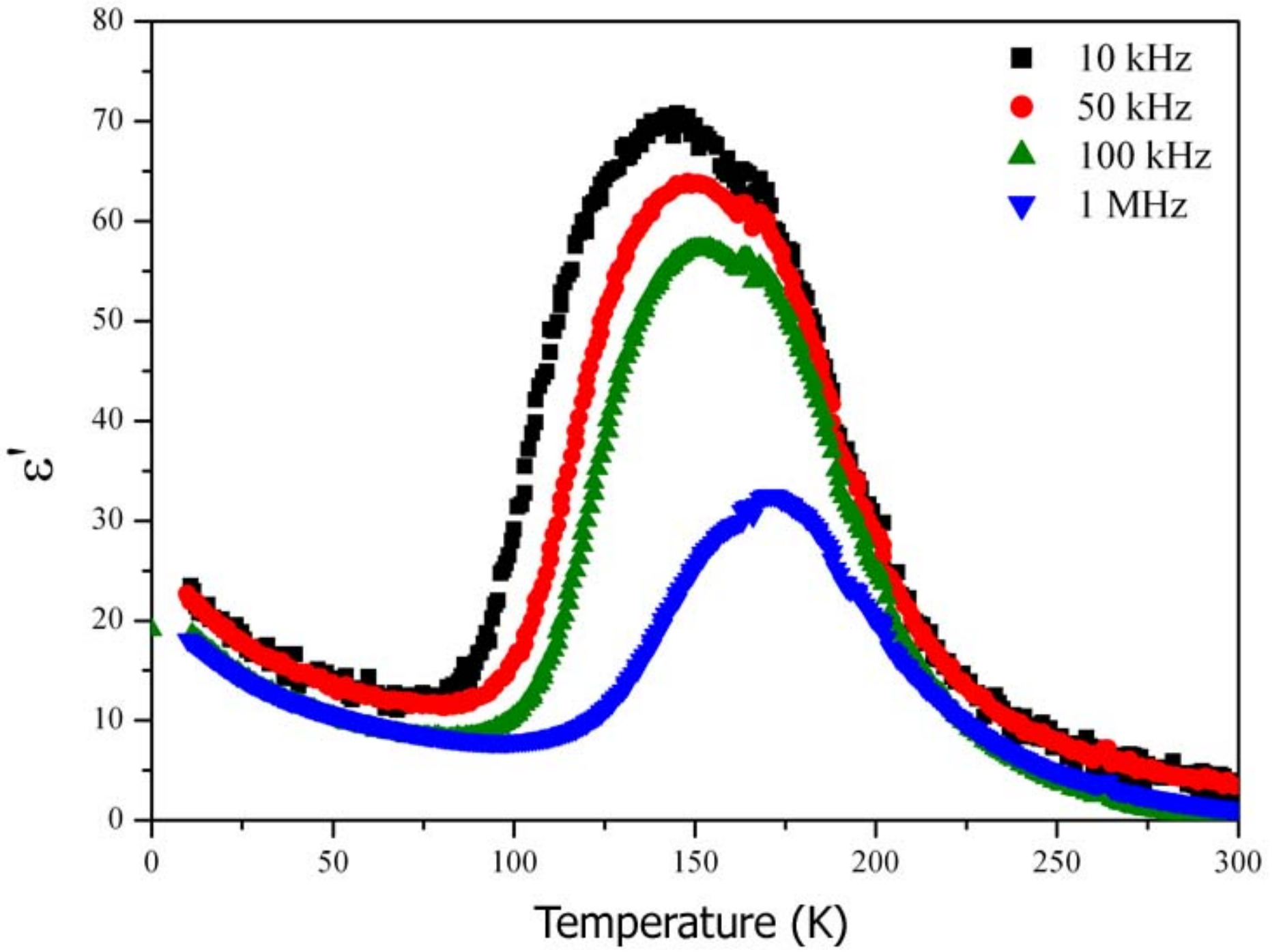

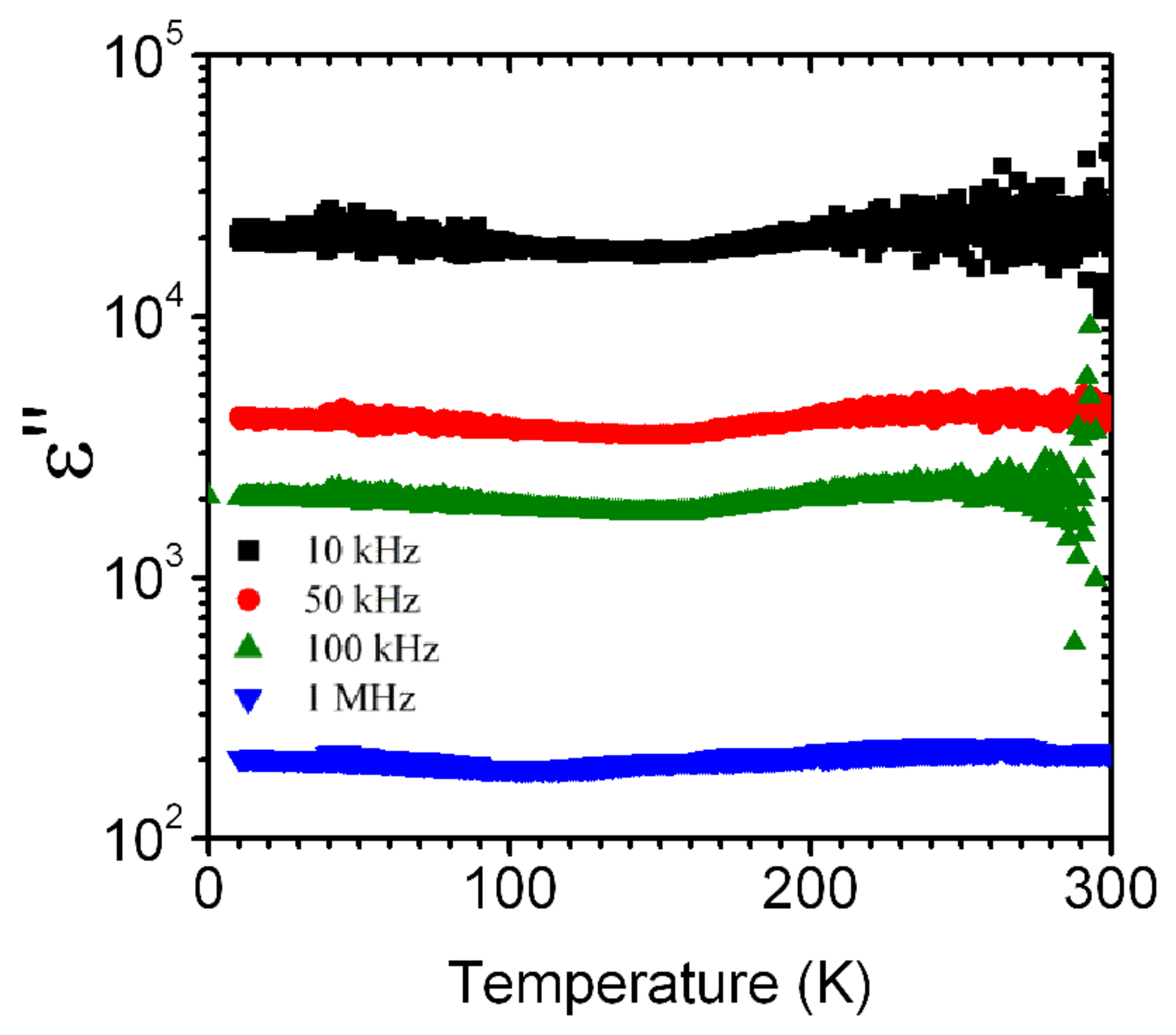